\documentclass{article}

\usepackage{microtype}
\usepackage{amssymb}
\usepackage{amsmath}
\usepackage{amsthm}
\usepackage{alltt}
\usepackage{color}
\usepackage{graphicx}
\usepackage{authblk}

\newtheorem{introtheorem}{Theorem}
\newtheorem{theorem}{Theorem}
\newtheorem{lemma}[theorem]{Lemma}

\newtheorem{corollary}[theorem]{Corollary}

\newenvironment{qedproof}{\begin{proof}}%
{\end{proof}}

\newcommand{\card}[1]{\left\lvert #1 \right\rvert}
\newcommand{\compl}[1]{\overline{#1}}

\newcommand{\var}{\mathtt{var}}
\newcommand{\sat}{\mathtt{sat}}
\newcommand{\lit}{\mathtt{lit}}
\newcommand{\cla}{\mathtt{cla}}
\newcommand{\bigoh}{\mathcal{O}}

\title{Solving {\sc MaxSAT} and {\sc \#SAT} on  structured CNF formulas}
\author{Sigve Hortemo S\ae{}ther}
\author{Jan Arne Telle}
\author{Martin Vatshelle}

\affil{Department of Informatics, University of Bergen, Norway}

\date{}%

\begin{document}
\maketitle{}

\begin{abstract}
In this paper we propose a structural parameter of CNF formulas and use it
to identify instances of weighted {\sc MaxSAT} and {\sc \#SAT} that can be solved in polynomial time.
Given a CNF formula we say that a set of clauses is precisely satisfiable if
there is some complete assignment satisfying these clauses only.
Let the $\mathtt{ps}$-value of the formula be the number of precisely satisfiable sets of clauses.
Applying the notion of branch decompositions to CNF
formulas and using $\mathtt{ps}$-value as cut function,
we define the $\mathtt{ps}$-width of a formula.
For a formula given with a decomposition of polynomial
$\mathtt{ps}$-width we show dynamic programming algorithms solving
weighted {\sc MaxSAT} and {\sc \#SAT} in polynomial time.  Combining
with results of 'Belmonte and Vatshelle, Graph classes with structured
neighborhoods and algorithmic applications, {\sc Theor. Comput. Sci.}
511: 54-65 (2013)' we get polynomial-time algorithms solving weighted
{\sc MaxSAT} and {\sc \#SAT} for some classes of structured CNF
formulas.  For example, we get $\bigoh(m^2(m + n)s)$ algorithms for
formulas $F$ of $m$ clauses and $n$ variables and size $s$, if $F$ has a
linear ordering of the variables and clauses such that for any
variable $x$ occurring in clause $C$, if $x$ appears before $C$ then
any variable between them also occurs in $C$, and if $C$ appears
before $x$ then $x$ occurs also in any clause between them.  Note that
the class of incidence graphs of such formulas do not have bounded
clique-width.
\end{abstract}

\section{Introduction}

Given a CNF formula, propositional model counting ({\sc \#SAT}) is the problem of computing the number of 
satisfying assignments, and maximum satisfiability ({\sc MaxSAT}) is the problem of determining the maximum 
number of clauses that can be satisfied by some assignment.
Both problems are significantly harder than simply deciding if a satisfying assignment exists.
{\sc \#SAT} is \#P-hard \cite{DBLP:books/fm/GareyJ79} even when restricted to Horn 2-CNF formulas, and to monotone 2-CNF formulas
\cite{DBLP:conf/aaai/Roth96}. 
{\sc MaxSAT} is NP-hard even when restricted to Horn 2-CNF formulas 
\cite{DBLP:journals/ipl/JaumardS87}, and
to 2-CNF formulas where each variable appears at most 3 times
\cite{DBLP:journals/ipl/RamanRR98}. 
Both problems become tractable 
under certain structural restrictions obtained by bounding width parameters of graphs associated with formulas,
see for example \cite{DBLP:journals/dam/FischerMR08,DBLP:journals/fuin/GanianHO13,DBLP:journals/jda/SamerS10,DBLP:conf/sat/Szeider03}.
The work we present here is inspired by the recent results of Paulusma et al \cite{DBLP:conf/stacs/PaulusmaSS13}
and
Slivovsky and Szeider 
\cite{DBLP:conf/isaac/SlivovskyS13} showing that
{\sc \#SAT} is solvable in polynomial time when the incidence graph $I(F)$ of the input formula $F$
has bounded modular treewidth, and more strongly, bounded symmetric clique-width.

We extend these results in several ways. We give algorithms for both {\sc \#SAT} and {\sc MaxSAT}, and also weighted {\sc MaxSAT},
finding the maximum weight of satisfiable clauses, given a set of weighted clauses.
We introduce the parameter $\mathtt{ps}$-width, and express the runtime of our algorithms 
as a function of $\mathtt{ps}$-width.

\setcounter{introtheorem}{2}
\begin{introtheorem}
  Given a formula $F$ over $n$ variables and $m$ clauses and of size $s$, and a decomposition of $F$ of
  $\mathtt{ps}$-width $k$, we solve 
  \textsc{\#SAT}, and weighted \textsc{MaxSAT} in time
  $\bigoh(k^3s(m + n))$.%
\end{introtheorem}
Thus, given a decomposition having a $\mathtt{ps}$-width $k$ that is {\em polynomially-bounded} in the number of variables $n$ and clauses $m$ of the formula, we get polynomial-time algorithms.
These are dynamic programming algorithms similar to the one given for {\sc \#SAT} in \cite{DBLP:conf/isaac/SlivovskyS13}, but we believe that
the $\mathtt{ps}$-width parameter is a better measure of its 'inherent runtime bottleneck'.
The essential combinatorial result enabling this improvement is Lemma \ref{lem:ps_leq_mimw} of this paper.
The algorithm of \cite{DBLP:conf/isaac/SlivovskyS13} solves \textsc{\#SAT} in time $(n+m)^{\bigoh(w)}$ for $w$ being the symmetric clique-width of the decomposition, and is thus a polynomial-time algorithm if given a decomposition with {\em constantly bounded} $w$. The result of Theorem \ref{theorem:SolvingSATonBranchDec} encompasses this, since we show via the concept of MIM-width \cite{Vatshelle}, that any formula with 
constantly bounded symmetric clique-width also has polynomially bounded $\mathtt{ps}$-width.

\begin{center}
  \begin{figure}[h]
    \includegraphics[scale=0.6]{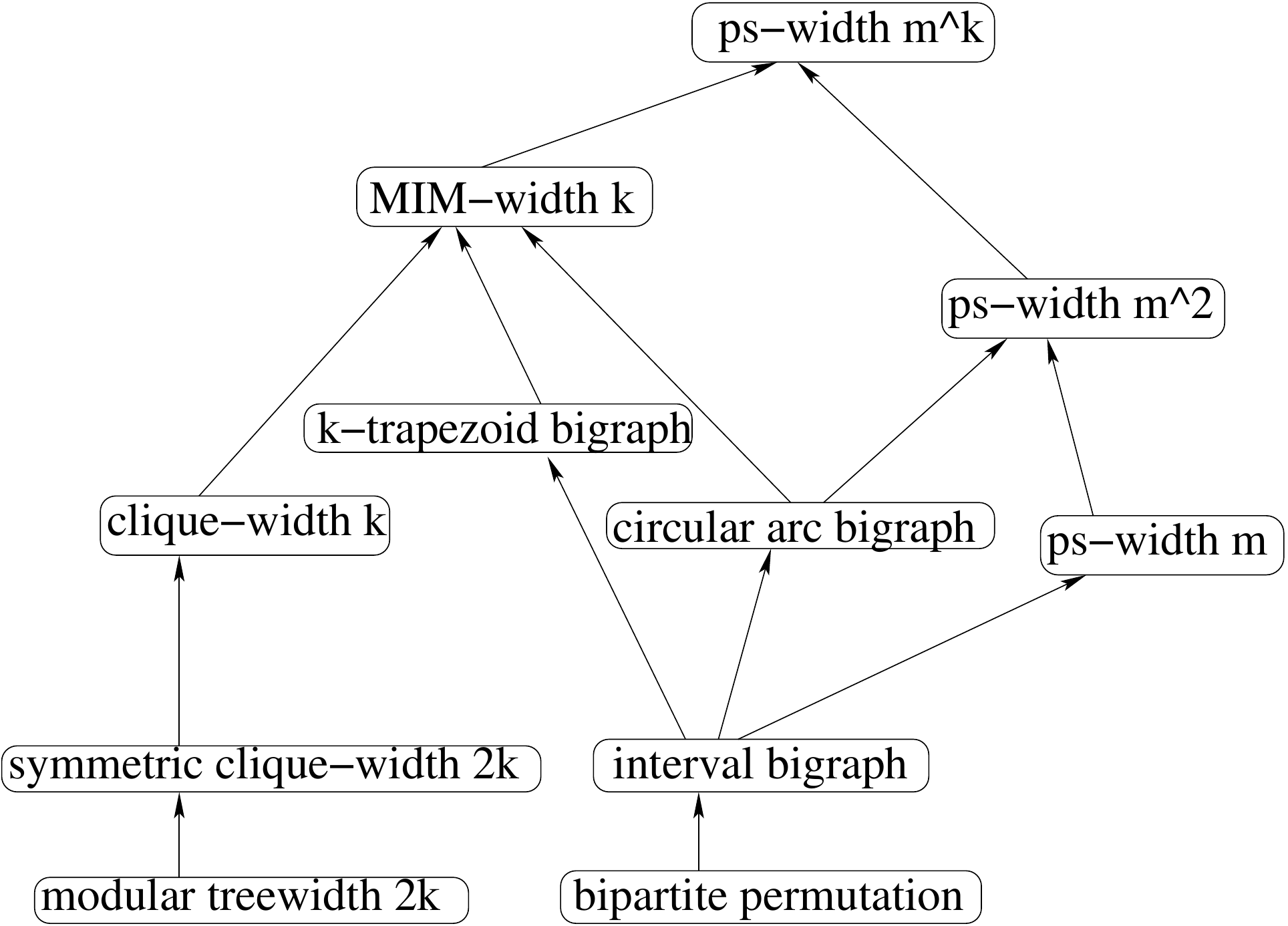}
 \caption{A hierarchy of structural parameters and classes of bipartite graphs, where $k$ is a constant and $F$ a CNF formula having $m$ clauses.
 An arc from $P$ to $Q$ means
 'any formula (or incidence graph of a formula) that has a decomposition of type $P$, also has a decomposition of type $Q$'.
 The lack of an arc means that no such relation holds, i.e. this is a Hasse diagram.}
\label{fig:6}
\end{figure}

\end{center}

We show that a relatively rich class of formulas, including classes of unbounded clique-width, have polynomially bounded 
$\mathtt{ps}$-width. This is shown using the concept of MIM-width of graphs, introduced in the thesis of Vatshelle \cite{Vatshelle}. See Figure \ref{fig:6}.
In particular, this holds for classes of formulas having incidence graphs that can be represented as intersection graphs of certain objects, like interval graphs \cite{DBLP:journals/tcs/BelmonteV13}.
We prove this also for bigraph bipartizations of these graphs, which are obtained by imposing a bipartition on the vertex set and keeping only edges between the partition classes. 
Some such bigraph bipartizations have been studied previously, in particular the interval bigraphs.
The interval bigraphs contain all bipartite permutation graphs, and these latter graphs have been shown to have unbounded clique-width \cite{DBLP:journals/arscom/BrandstadtL03}. 

By combining an alternative definition of interval bigraphs \cite{DBLP:journals/jgt/HellH04} with a fast recognition algorithm \cite{DBLP:journals/dam/Muller97,DBLP:journals/corr/abs-1211-2662} we arrive at the following.
Say that a CNF formula $F$ has an interval ordering if there exists a linear ordering of variables and clauses such that for any variable $x$ occurring in clause $C$, if $x$ appears before $C$
then any variable between them also occurs in $C$, and if $C$ appears before $x$
then $x$ occurs also in any clause between them.

\setcounter{introtheorem}{9}
\begin{introtheorem}
 Given a CNF formula
  $F$  over $n$ variables and $m$ clauses and of size $s$, we can in time $\bigoh((m + n)s)$
  decide if $F$ has an interval ordering (yes iff $I(F)$ is an
  interval bigraph), and if yes we solve \textsc{\#SAT} and weighted
  \textsc{MaxSAT} with a runtime of
  $\bigoh(m^2(m + n)s)$.
\end{introtheorem}

The algorithms of Theorem \ref{thm:intbigraph_algorithm} may be of interest for practical applications, as there are no big hidden constants in the runtimes.

Our paper is organized as follows. In Section 2 we give formal definitions.
We will be using a type of decomposition that originates in the theory of graphs and matroids where it is known as branch decomposition, see~\cite{GGW02,GMX}. 
The standard approach is to apply this type of decomposition to the incidence graph of a formula, and 
evaluate its width using as cut function a graph parameter, as done in \cite{DBLP:conf/isaac/SlivovskyS13}. 
The cut function we will use is not a graph parameter, but rather the $\mathtt{ps}$-value of a formula, being the number of distinct subsets of clauses that are satisfied by some complete assignment. We thus prefer to apply the decomposition directly to the formula and not to its incidence graph, although the translation between the two will be straightforward.
We define cuts of formulas and $\mathtt{ps}$-width of a formula. Note that a formula can have $\mathtt{ps}$-value exponential and $\mathtt{ps}$-width polynomial.
In Section 3 we present dynamic programming algorithms that given a formula and a decomposition solves \textsc{\#SAT} and
 weighted \textsc{MaxSAT}, proving Theorem \ref{theorem:SolvingSATonBranchDec}. In Section 4 we investigate classes of formulas having decompositions of low $\mathtt{ps}$-width, basically proving the correctness of the hierarchy presented in Figure \ref{fig:6}.
 In Section 5 we consider formulas having an interval ordering and prove Theorem \ref{thm:intbigraph_algorithm}.
 We end in Section 5 with some open problems.

\section{Framework}

A \emph{literal} is a propositional \emph{variable} or a negated
variable, $x$ or $\neg x$, a \emph{clause} is a set of literals, and a
\emph{formula} is a multiset of clauses. For a formula $F$, $\cla(F)$
denotes the clauses in $F$. For a clause $C$, $\lit(C)$ denotes the
set of literals in $C$ and $\var(C)$ denotes the variables of the
literals in $\lit(C)$. For a set $S$ of variables and clauses, $
\var(S)$ denotes the variables of $S$ and $\cla(S)$ denotes the
clauses.  For a formula $F$, $\var(F)$ denotes the union $\bigcup_{C
  \in \cla(F)} \var(C)$.  For a set $X$ of variables, an
\emph{assignment} of $X$ is a function $\tau : X \to \{0,1\}$.  For a
literal $\ell$, we define $\tau(\ell)$ to be $1 - \tau(\var(\ell))$ if
$\ell$ is a negated variable ($\ell = \neg x$ for some variable $x$)
and to be $\tau(\var)$ otherwise ($\ell = x$ for some variable $x$).
A clause $C$ is said to be \emph{satisfied} by an assignment $\tau$ if
there exists at least one literal $\ell \in \lit(C)$ so that
$\tau(\ell) = 1$. All clauses an assignment $\tau$ do not satisfy are
said to be \emph{unsatisfied} by $\tau$. We notice that this means an
empty clause will be unsatisfied by all assignments. A formula is
satisfied by an assignment $\tau$ if $\tau$ satisfies all clauses in
$\cla(F)$.
 
The problem \textsc{\#SAT}, given a formula $F$, asks how many
distinct assignments of $\var(F)$ satisfy $F$.  The optimization
problem weighted \textsc{MaxSAT}, given a formula $F$ and weight
function $w : \cla(F) \to \mathbb{N}$, asks what assignment $\tau$ of
$\var(F)$ maximizes $\sum_C w(C)$ for all $C \in \cla(F)$ satisfied by
$\tau$. The problem \textsc{MaxSAT} is weighted \textsc{MaxSAT} where
all clauses have weight one.  When given a CNF formula $F$, we use $s$
to denote the size of $F$. More precisely, the size of $F$ is $s =
\card{\cla(F)} + \sum_{C \in \cla(F)} \card{\lit(C)}$.  For weighted
\textsc{MaxSAT}, we assume the sum of all the weights are at most
$2^{O(\cla{F})}$, and thus we can do summation on the weights in time
linear in $\cla{F}$.

For a set $A$, with elements from a universe $U$ we denote by $\compl{A}$ the elements
in $U \setminus A$, as the universe is usually given by the context.

\subsection{Cut of a formula}

In this paper, we will solve \textsc{MaxSAT} and \textsc{\#SAT} by the
use of dynamic programming. We will be using a divide and conquer
technique where we solve the problem on smaller subformulas of the
original formula $F$ and then combine the solutions to each of these
smaller formulas to form a solution to the entire formula $F$. 
Note however, that the solutions found for a subformula will depend on the 
interaction between the subformula and the remainder of the formula.
We use the following notation for subformulas.

For a clause $C$ and set $X$ of variables, by $C \vert_{X}$ we denote
the clause $\{\ell \in C: \var(\ell) \in X\}$. We say $C \vert_{X}$ is
the clause $C$ \emph{induced} by $X$. For a formula $F$ and subsets
$\mathcal{C} \subseteq \cla(F)$ and $X \subseteq \var(F)$, we say the
subformula $F_{\mathcal{C},X}$ of $F$ \emph{induced} by $\mathcal{C}$
and $X$ is the formula consisting of the clauses $\{C_i
\vert_{X} : C_i \in \mathcal{C}\}$. That is, $F_{\mathcal{C}, X}$ is
the formula we get by removing all clauses not in $\mathcal{C}$
followed by removing each literal that consists of a variable not in
$X$.
As with a clause, for an assignment $\tau$ over a set $X$ of
variables, we say the assignment $\tau$ \emph{induced} by $X'
\subseteq X$ is the assignment $\tau \vert_{X'}$ where the domain is
restricted to $X'$. %

For a formula $F$ and sets $\mathcal{C} \subseteq \cla(F)$, $X
\subseteq \var(F)$, and $S = \mathcal{C} \cup X$, we call $S$ a
\emph{cut} of $F$ and note that it breaks $F$ into four subformulas $F_{\mathcal{C},X}$,
$F_{\compl{\mathcal{C}},X}$, $F_{\mathcal{C},\compl{X}},$ and
$F_{\compl{\mathcal{C}}, \compl X}.$ See Figure \ref{fig:CutOfFormula}. 
One important fact we may observe
from this definition is that a clause $C$ in $F$ is satisfied by an
assignment $\tau$ of $\var(F)$, if and only if $C$ (induced by $X$ or
$\compl X$) is satisfied by $\tau$ in at least one of the formulas of
any cut of $F$.

\subsection{Precisely satisfiable sets and \texttt{ps}-value of a formula}

For a formula $F$ and assignment $\tau$ of all the variables in
$\var(F)$, we denote by $\sat(F, \tau)$ the set $\mathcal{C} \subseteq
\cla(F)$ so that each clause in $\mathcal{C}$ is satisfied by $\tau$,
and each clause not in $\mathcal{C}$ is unsatisfied by $\tau$. If for
a set $\mathcal{C} \subseteq \cla(F)$ we have $\sat(F, \tau) =
\mathcal{C}$ for some $\tau$ over $\var(F)$, we say $\mathcal{C}$ is
\emph{precisely satisfiable} in $F$.  We denote by $\mathtt{PS}(F)$ the family
of all precisely satisfiable sets in $F$. That is,
\[
  \mathtt{PS}(F) = %
    \{
      \sat(F,\tau)   : 
      \text{$\tau$ is an assignment of $\var(F)$}
    \}.
\]

The cardinality of this set,  $\mathtt{PS}(F)$, is referred to as the $\mathtt{ps}$-value of $F$. 

\subsection{The \texttt{ps}-width of a formula}

We define a \emph{branch decomposition} of a formula $F$ to be a pair
$(T, \delta)$ where $T$ is a rooted binary tree and $\delta$ is a
bijective function from the leaves of $T$ to the clauses and variables
of $F$. If all the non-leaf nodes (also referred to as \emph{internal}
nodes) of $T$ induce a path, we say that $(T, \delta)$ is a
\emph{linear} branch decomposition.  For a non-leaf node $v$ of $T$,
we denote by $\delta(v)$ the set $\{\delta(l): \text{\(l\) is a leaf
  in the subtree rooted in \(v\)}\}$. Based on this, we say that the
decomposition $(T, \delta)$ of formula $F$ induces certain cuts of
$F$, namely the cuts defined by $\delta(v)$ for each node $v$ in $T$.

For a formula $F$ and branch decomposition $(T, \delta)$, for each
node $v$ in $T$, by $F_v$ we denote the formula induced by the
clauses in $\cla(F) \setminus \delta(v)$ and the variables in
$\delta(v)$, and by $F_{\compl v}$ we denote the formula on the
complement sets; i.e. the clauses in $\delta(v)$ and the variables in
$\var(F) \setminus \delta(v)$.
In other words, if $\delta(v)= \mathcal{C} \cup X$
with $\mathcal{C} \subseteq \cla(F)$ and $X
\subseteq \var(F)$ then  
$F_v=F_{\compl{\mathcal{C}},X}$ and $F_{\compl v}=F_{\mathcal{C},\compl{X}}$.
We define the 
\emph{$\mathtt{ps}$-value} of the cut $\delta(v)$ to be
\[ \mathtt{ps}(\delta(v)) = \max\{|PS(F_v)|,|PS(F_{\compl v})|\}
 \]
We define the
\emph{$\mathtt{ps}$-width} of a branch decomposition to be 
\[ \mathtt{psw}(T, \delta) = \max\{
\mathtt{ps}(\delta(v)): v \text{ is a node of } T\}
 \]
We define the \emph{$\mathtt{ps}$-width} of a formula
$F$ to be
\[ \mathtt{psw}(F) = \min \{\mathtt{psw}(T, \delta): (T, \delta) \text{ is a branch decompositions of }F\}
 \]

\begin{figure}[ht!]
  \centering
  \includegraphics[scale=1.2]{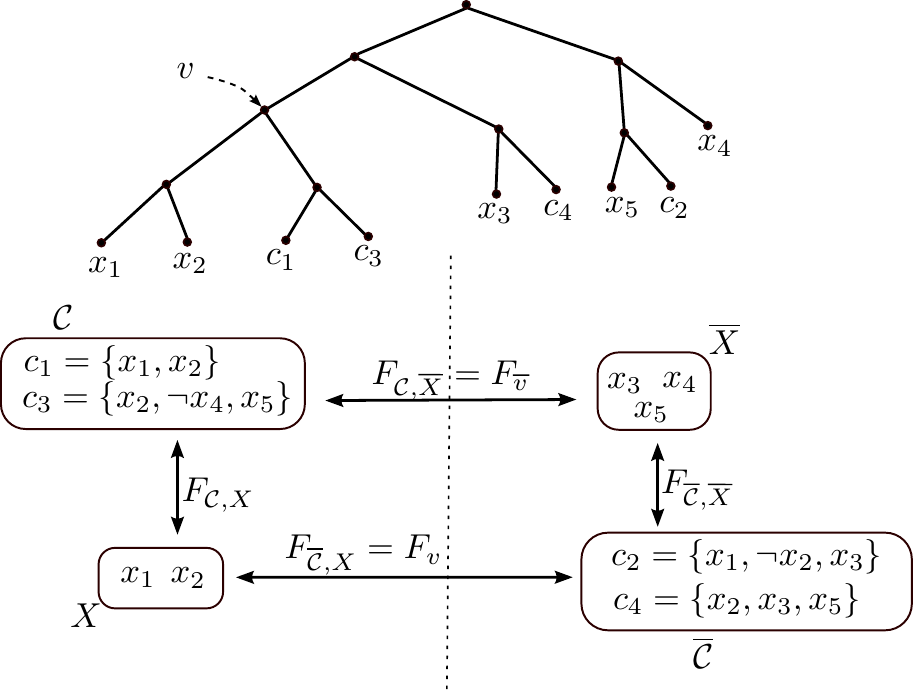}
  \caption{On top is a branch decomposition of a formula $F$ with $\var(F)=\{x_1,x_2,x_3,x_4,x_5\}$ and
  the 4 clauses $\cla(F)=\{c_1,c_2,c_3,c_4\}$ as given in the boxes.
  The node $v$ of the tree defines the cut 
   $\delta(v)= \mathcal{C} \uplus X$ where $\mathcal{C}=\{c_1,c_3\}$ and
   $X=\{x_1,x_2\}$.
   On the bottom is an illustration of the 4 subformulas defined by this cut.  
    For example,  $F_{\compl{\mathcal{C}}, X} =
    \{\{x_1, \neg{x_2}\},\{x_2\}\}$ and $F_{\mathcal{C}, \compl X} =
    \{\emptyset,\{\neg{x_4}, x_5\}\}$.
  We have $F_v=F_{\compl{\mathcal{C}}, X}$ and $F_{\compl
      v}=F_{\mathcal{C}, \compl X}$ with perfectly satisfiable sets of clauses 
      $\mathtt{PS}(F_v)=\{\{c_2\}, \{c_4\}, \{c_2, c_4\}\}$ and $\mathtt{PS}(F_{\compl
      v})=\{\emptyset, \{c_3\}\}$  and the $\mathtt{ps}$-value of
 this cut is  $\mathtt{ps}(\delta(v)) = \max\{|PS(F_v)|,|PS(F_{\compl v})|\}=3  $.}
\label{fig:CutOfFormula}
\end{figure}

Note that the $\mathtt{ps}$-value of a cut is a symmetric function. That is, the
$\mathtt{ps}$-value of cut $S$ equals the $\mathtt{ps}$-value of the cut $\compl
S$. See Figure \ref{fig:CutOfFormula} for an example.

\section{Dynamic programming for {\sc MaxSAT} and {\sc \#SAT} }

Given a branch decomposition $(T, \delta)$ of a CNF formula $F$ over $n$ variables and $m$ clauses and of size $s$, we
will give algorithms that solve  {\sc MaxSAT} and {\sc \#SAT} on $F$
in time $\bigoh(\mathtt{psw}(T, \delta)^3s(m + n))$.

In a pre-processing step we will need the following which, for each node $v$ in $T$
computes the sets $\mathtt{PS}(F_v)$ and $\mathtt{PS}(F_{\compl
  v})$. 
\begin{theorem} \label{thm:constr.PSsets} Given a CNF formula $F$ of $n$ variables and $m$ clauses with
  a branch decomposition $(T, \delta)$ of $\mathtt{ps}$-width $k$, we
  can in time $\bigoh(k^2\log(k)m(m+n))$ compute
  the sets $\mathtt{PS}(F_v)$ and $\mathtt{PS}(F_{\compl v})$ for each
  $v$ in $T$.
\end{theorem}

\begin{qedproof}
  We notice that for a node $v$ in $T$ with children $c_1$ and $c_2$,
  we can express $\mathtt{PS}(F_v)$ as
  \[
  \mathtt{PS}(F_v) =%
  \left\{%
    (C_1 \cup C_2) \cap \cla(F_v) %
    : %
    C_1 \in \mathtt{PS}(F_{c_1}), %
    C_2 \in \mathtt{PS}(F_{c_2}) %
  \right\}.
  \]
  Similarly, for sibling $s$ and parent $p$ of $v$ in $T$, the set
  $\mathtt{PS}(F_{\compl v})$ can be expressed as
  \[ 
  \mathtt{PS}(F_{\compl v}) = %
  \left\{ %
    (C_p \cup C_s) \cap \cla(F_{\compl v}) %
    : %
    C_p \in \mathtt{PS}(F_{\compl p}), %
    C_s \in \mathtt{PS}(F_{s}) %
  \right\}.
  \]

  By transforming these recursive expressions into a dynamic
  programming algorithm, as done in Procedure~1 and Procedure~2 below,
  we are able to calculate all the desired sets as long as we can
  compute the sets for the base cases $\mathtt{PS}(F_l)$ when $l$ is a
  leaf of $T$, and $\mathtt{PS}(F_{\compl r})$ for the root $r$ of
  $T$. However, these formulas contain at most one variable, and thus
  we can easily construct their set of specific satisfied clauses in
  linear amount of time for each of the formulas. For the rest of the
  formulas, we construct the formulas using Procedure~1 and
  Procedure~2. As there are at most twice as many nodes in $T$ as
  there are clauses and variables in $F$, the procedures will run at
  most $\bigoh(|\cla(F)| + |\var(F)|)$ times. In each run of the
  algorithms, we iterate through at most $k^2$ pairs of precisely
  satisfiable sets, and do a constant number of set operations that
  might take $\bigoh(|\cla(F)|)$ time each. Then we sort the list of
  at most $k^2$ sets of clauses. When we sort, we can expect the
  runtime of comparing two elements to spend time linear in $|\cla(F)|$,
  so the total runtime for sorting $L$ and deleting duplicates takes
  at most $\bigoh(k^2\log(k)|\cla(F)|)$ time. This results in a total runtime of
  $\bigoh(k^2\log(k)|\cla(F)|(|\cla(F)|+|\var(F)|))$ for all the nodes of $T$ combined.
\end{qedproof}

\begin{figure}[h!]
  \center
  \begin{tabular}{|l|}
    \hline
    Procedure 1: Generating $\mathtt{PS}(F_v)$\\
    \hline \hline
    \begin{minipage}{1.0\linewidth}
    \begin{tabbing}
      \=\textbf{output:} \=\kill
      \>\textbf{input:}\>$\mathtt{PS}(F_{c_1})$ and
       $\mathtt{PS}(F_{c_2})$ for children $c_1$ and 
       $c_2$ of $v$\\ 
       \> \>in branch decomposition \\
      \>\textbf{output:}\>$\mathtt{PS}(F_v)$
    \end{tabbing}
  \end{minipage}\\\hline \hline
  \begin{minipage}{1.0\linewidth}
    \vspace{2pt}
    \begin{tabbing}
      \=xx\=xx\=xx\=xx\=xx\=\kill
      \> $L \leftarrow$ empty list of precisely satisfiable clause-sets\\
      \> \textbf{for} each $(C_1, C_2) \in
      \mathtt{PS}(F_{c_1}) \times \mathtt{PS}(F_{c_2})$
      \textbf{do}\\
      \> \> add $(C_1 \cup C_2) \setminus \cla(\delta(v))$
      to $L$\\
      \> sort $L$ lexicographically by what clauses each
      element contains\\
      \> remove duplicates in $L$ by looking only at consecutive
      elements\\
      \> \textbf{return} $L$
    \end{tabbing}    
  \end{minipage}\\
 \hline
\end{tabular}
\end{figure}

\begin{figure}[h!] 
  \center
  \begin{tabular}{|l|}
    \hline
    Procedure 2: Generating $\mathtt{PS}(F_{\compl v})$\\
    \hline \hline
    \begin{minipage}{1.0\linewidth}
    \begin{tabbing}
      \=\textbf{output:} \=\kill
      \>\textbf{input:}\>$\mathtt{PS}(F_{s})$ and
       $\mathtt{PS}(F_{\compl p})$ for sibling $s$ and 
       parent $p$ of $v$\\
       \> \> in branch decomposition \\
      \>\textbf{output:}\>$\mathtt{PS}(F_{\compl v})$
    \end{tabbing}
  \end{minipage}\\\hline \hline
  \begin{minipage}{1.0\linewidth}
    \vspace{2pt}
    \begin{tabbing}
      \=xx\=xx\=xx\=xx\=xx\=\kill
      \> $L \leftarrow$ empty list of precisely satisfiable clause-sets\\
      \> \textbf{for} each $(C_s, C_p) \in
      \mathtt{PS}(F_{s}) \times \mathtt{PS}(F_{\compl p})$
      \textbf{do}\\
      \> \> add $(C_s \cup C_p) \setminus \cla(\delta(v))$
      to $L$\\
      \> sort $L$ lexicographically by what clauses each
      element contains\\
      \> remove duplicates in $L$ by looking only at consecutive
      elements\\
      \> \textbf{return} $L$
    \end{tabbing}    
  \end{minipage}\\
 \hline
\end{tabular}
\end{figure}

We first give the algorithm for {\sc MaxSAT} and then briefly describe the
changes necessary for solving 
weighted {\sc MaxSAT} and {\sc \#SAT}.

Our algorithm relies on the following binary relation,
$\leq$, on assignments $\tau$ and $\tau'$ related to a cut $S =
\mathcal{C} \cup X$ with
$\mathcal{C} \subseteq \cla(F)$, $X \subseteq \var(F)$. For $\mathcal{C}' \in \mathtt{PS}(F_{\mathcal{C},
  \compl X})$ we define $\tau' \leq_{S}^{\mathcal{C}'} \tau$ if it holds that
$\card{ \sat(F, \tau') \setminus \mathcal{C}' } \leq \card{ \sat(F,
  \tau) \setminus \mathcal{C}' }$. 
  Note that for each cut $S =
\mathcal{C} \cup X$ and each $\mathcal{C}' \in \mathtt{PS}(F_{\mathcal{C},
  \compl X})$ this gives a total preorder (transitive, reflexive and total) on assignments.
  The largest elements of this total preorder will be important for our algorithm, as they satisfy the maximum number of clauses under the given restrictions.
  
Given $(T,\delta)$ of a formula $F$ our
dynamic programming algorithm for {\sc MaxSAT} will generate, for each node $v$ in $T$, a
table $\mathtt{Tab_v}$ indexed by pairs of $\mathtt{PS}(F_v) \times
\mathtt{PS}(F_{\compl v})$.
For precisely satisfiable sets
$C_v \in \mathtt{PS}(F_v)$ and $C_{\compl v} \in \mathtt{PS}(F_{\compl v})$ the
contents of the table at this index $\mathtt{Tab_v}(C_v, C_{\compl v})$ should be an assignment 
$\tau:\var(\delta(v)) \to \{0,1\}$ satisfying the following constraint:
\begin{equation}
\label{eq}
\begin{aligned} 
\mathtt{Tab_v}(C_v, C_{\compl v})= \tau &\text{ such that } 
 \sat(F_v, \tau) = C_v
 \text{ and } \tau' \leq_{\delta(v)}^{C_{\compl v}} \tau \text{ for any } \\
 &\tau':\var(\delta(v)) \to \{0,1\} \text{ having } 
 \sat(F_v, \tau') = C_v
\end{aligned}
\end{equation} 
Let us give some intuition for this constraint. 
Our algorithm uses the technique of 'expectation from the outside' introduced in \cite{DBLP:journals/dam/Bui-XuanTV10,BTV}.
The partial assignment $\tau$ to variables
in $\var(\delta(v))$ 
stored at $\mathtt{Tab_v}(C_v, C_{\compl v})$ 
will be combined with partial assignments to variables in $\var(F) \setminus \var(\delta(v))$ satisfying $C_{\compl v}$.
These latter partial assignments 
constitute 'the expectation from the outside'.
Constraint (\ref{eq}) implies that $\tau$, being a largest element of the total preorder, will be a best combination with 
this expectation from the outside since it satisfies the maximum number of remaining clauses.

By bottom-up dynamic programming along the tree $T$ we compute the tables of each node of $T$.
For a leaf $l$ in $T$,  generating $\mathtt{Tab}_l$ can be done easily in linear time
since the formula $F_v$ contains at most one
variable.  For an internal node $v$ of $T$, with children $c_1,c_2$, we compute $\mathtt{Tab_v}$ by the
algorithm described in Procedure~3.
There are 3 tables involved in this update, one at each child and one at the parent.
A pair of entries, one from each child table, may lead to an update of an entry in the parent table. Each table entry is indexed by a pair, thus there are 6 indices involved in a single potential update.
 A clever trick first introduced in \cite{BTV} allows us to loop over triples of indices and for each triple 
 compute the remaining 3 indices forming the 6-tuple involved in the update, thereby reducing the runtime.

\begin{center} \label{proc:constrDPtable}
  \begin{tabular}{|l|}
    \hline
    Procedure 3: Computing $\mathtt{Tab_{v}}$ for inner node $v$ with
    children $c_1, c_2$\\
    \hline \hline
    \begin{minipage}{1.0\linewidth}
    \begin{tabbing}
      
      \=\textbf{output:} \=\kill
      \>\textbf{input:} \> $\mathtt{Tab_{c_1}}$, $\mathtt{Tab_{c_2}}$\\
      \>\textbf{output:}\>$\mathtt{Tab_{v}}$
    \end{tabbing}
  \end{minipage}\\\hline \hline
  \begin{minipage}{1.0\linewidth}
    \vspace{2pt}
    \begin{tabbing}
      xxx\=xx\=xx\=xx\=xx\=xx\=\kill

1.      \>initialize $\mathtt{Tab_v} : 
                      \mathtt{PS}(F_{v}) \times \mathtt{PS}(F_{\compl v}) 
                      \to \{\mathtt{unassigned}\}$  \emph{// dummy entries}\\
2.      \> \textbf{for} each $(C_{c_1}, C_{c_2}, C_{\compl v}) \in 
                               \mathtt{PS}(F_{c_1})         \times  
                               \mathtt{PS}(F_{c_2})         \times 
                               \mathtt{PS}(F_{\compl v})$ 
         \textbf{do} \\

3.      \> \> \> \> $C_{\compl {c_1}} \leftarrow (C_{c_2} \cup {C_{\compl v}}) \cap \delta(c_1)$\\
4.      \> \> \> \> $C_{\compl {c_2}} \leftarrow (C_{c_1} \cup {C_{\compl v}}) \cap \delta(c_2)$\\

5.      \> \> \> \> $C_v \leftarrow (C_{c_1} \cup C_{c_2})  \setminus \delta(v)$\\
6.      \> \> \> \> $\tau \leftarrow \mathtt{Tab_{c_1}}(C_{c_1}, C_{\compl {c_1}}) \uplus \mathtt{Tab_{c_2}}(C_{c_2}, C_{\compl {c_2}})$\\
7.      \> \> \> \> $\tau' \leftarrow \mathtt{Tab_v}(C_v, {C_{\compl v}})$ \\
8.      \> \> \> \> \textbf{if} $\tau' = \mathtt{unassigned}$ or $\tau \geq_{\delta(v)}^{{C_{\compl v}}} \tau'$  \textbf{then} 
      $\mathtt{Tab_v}(C_v, {C_{\compl v}}) \leftarrow \tau$\\
9.      \> \textbf{return} $\mathtt{Tab_v}$
    \end{tabbing}    
  \end{minipage}\\
 \hline
\end{tabular}
\end{center}

\begin{lemma} \label{lemma:tableRuntime} For a CNF formula $F$ of size
  $s$ and an inner node $v$, of a branch decomposition $(T, \delta)$
  of $\mathtt{ps}$-width $k$, Procedure~3 computes $\mathtt{Tab_v}$
  satisfying Constraint (\ref{eq}) in time $\bigoh(k^3s)$.
\end{lemma}

\begin{qedproof}
 We assume $\mathtt{Tab}_{c_1}$ and
  $\mathtt{Tab}_{c_2}$ satisfy Constraint (\ref{eq}). 
  Procedure~3 loops over all triples $(C_{c_1}, C_{c_2}, C_{\compl v}) \in 
                               \mathtt{PS}(F_{c_1})         \times  
                               \mathtt{PS}(F_{c_2})         \times 
                               \mathtt{PS}(F_{\compl v})$.                               
From the  definition of   $\mathtt{ps}$-width of  $(T, \delta)$ there are at most $k^3$ such triples.
Each operation inside an iteration of the loop take $\bigoh(s)$ time and there is a constant number of such operations.
Thus the runtime is $\bigoh(k^3s)$.

  To show that
  the output $\mathtt{Tab}_v$ of Procedure~3 satisfies Constraint (\ref{eq}), we will prove that for any $C \in
  \mathtt{PS}(F_v)$ and $C' \in \mathtt{PS}(F_{\compl v})$ the value
  of $\mathtt{Tab}_v(C, C')$ satisfies Constraint (\ref{eq}). That is, we will assure that
  the content of $\mathtt{Tab}_v(C, C')$ is an assignment $\tau$ so
  that $\sat(F_v, \tau) = C$ and for all other assignments $\tau'$
  over $\var(\delta(v))$ so that $\sat(F_v, \tau') = C$, we have $\tau'
  \leq_{\delta(v)}^{C'} \tau$.

  Let us assume for contradiction, that $\mathtt{Tab}_v(C, C')$
  contains an assignment $\tau$ but there exists an assignment $\tau'$
  over $\var(\delta(v))$ so that $\sat(F_v, \tau') = C$, and we do not have
  $\tau' \leq_{\delta(v)}^{C'} \tau$.  As $\tau$ is put into
  $\mathtt{Tab}_v(C, C')$ only if it is an assignment over $\var(\delta(v))$
  and $\sat(F_v, \tau) = C$. So, what we need to show to prove that
  $\mathtt{Tab}_v$ is correct is that in fact $\tau'
  \leq_{\delta(v)}^{C'} \tau$:

  First, we notice that $\tau'$ consist of assignments $\tau_1' =
  \tau' \vert_{\var(\delta(c_1))}$ and $\tau_2' = \tau'
  \vert_{\var(\delta(c_2))}$ where $\tau_1'$ is over the variables in
  $\var(\delta(c_1))$ and $\tau_2'$ is over $\var(\delta(c_2))$. Let $C_1 =
  \sat(F_{c_1}, \tau_1')$ and $C_2 = \sat(F_{c_2}, \tau_2')$ and let
  $C_1' = (C_2 \cup C') \cap \delta(c_1)$ and $C_2' = (C_1 \cup C')
  \cap \delta(c_2)$. By how $\mathtt{Tab}_{c_1}$ and
  $\mathtt{Tab}_{c_2}$ is defined, we know for the assignment $\tau_1$
  in $\mathtt{Tab}_{c_1}(C_1, C_1')$ and $\tau_2$ in
  $\mathtt{Tab}_{c_2}(C_2, C_2')$, we have $\tau_1' \leq_{\delta(c_1)}^{C_1'} \tau_1$ and
  $\tau_2' \leq_{\delta(c_2)}^{C_2'} \tau_2$. 
  From our definition of the total preorder $\leq$ for assignments,
  we can deduce that $\tau_1' \uplus \tau_2' \leq_{\delta(v)}^{C'}
  \tau_1 \uplus \tau_2$;
  \begin{align*}
  &\card{ \sat(F_v, \tau_1' \uplus \tau_2') \setminus C'} \\
              =& 
                   \card{ \sat(F_{c_1}, \tau_1') \setminus C_1'} - \card{C_1 \cap C_2'}
                +  \card{ \sat(F_{c_2}, \tau_2') \setminus C_2'} - \card{C_2 \cap C_1'}\\
              \leq&
                   \card{ \sat(F_{c_1}, \tau_1) \setminus C_1'} - \card{C_1 \cap C_2'}
                +  \card{ \sat(F_{c_2}, \tau_2) \setminus C_2'} - \card{C_2 \cap C_1'}\\
              =&
  \card{ \sat(F_v, \tau_1 \uplus \tau_2) \setminus C'}.
  \end{align*}
  However, since $\tau_1 \uplus \tau_2$ at the iteration of the
  triple $(C_1, C_2, C')$ in fact is considered by the algorithm to
  be set as $\mathtt{Tab}_v(C, C')$, it must be the case that $\tau_1
  \uplus \tau_2 \leq_{\delta(v)}^{C'} \tau$. As
  $\leq_{\delta(v)}^{C'}$ clearly is a transitive relation, we
  conclude that $\tau' \leq_{\delta(v)}^{C'} \tau$.
\end{qedproof}

\begin{theorem} \label{theorem:SolvingSATonBranchDec} Given a formula
  $F$ over $n$ variables and $m$ clauses and of size $s$, and a branch decomposition $(T, \delta)$ of $F$ of
  $\mathtt{ps}$-width $k$, we solve \textsc{MaxSAT}, \textsc{\#SAT},
  and weighted \textsc{MaxSAT} in time $\bigoh(k^3s(m + n))$.
\end{theorem}

\begin{qedproof}
  To solve \textsc{MaxSAT}, we first compute $\mathtt{Tab}_r$ for the
  root node $r$ of $T$. This requires that we first compute
  $\mathtt{PS}(F_v)$ and $\mathtt{PS}(F_{\compl v})$ for all nodes $v$
  of $T$, and then, in a bottom up manner, compute $\mathtt{Tab}_v$
  for each of the $\bigoh(m + n)$ nodes in $T$.  
  The former part we can do in $\bigoh(k^3s(m + n))$ time by
  Theorem~\ref{thm:constr.PSsets}, and the latter part we do in the
  same amount of time by Lemma~\ref{lemma:tableRuntime}.

  At the root $r$ of $T$ we have $\delta(r)= \var(F) \cup \cla(F)$.
  Thus $F_r = \emptyset$ and $F_{\compl r}$ contains only empty clauses, so that
  $PS(F_r) \times PS(F_{\compl r})$ contains only $(\emptyset, \emptyset)$.
  By Constraint (\ref{eq})
  and the definition of the $\leq$ total preorder on assignments, the assignment 
  $\tau$ stored in   $\mathtt{Tab}_r(\emptyset,
  \emptyset)$ is an assignment of $\var(F)$ maximizing $\card{ \sat(F, \tau)}$, the
  number of clauses satisfied, and hence is a solution to
  \textsc{MaxSAT}.

  For a weight function $w : \cla(F) \rightarrow \mathbb{N}$, by
  redefining $\tau_1 \leq_{A}^{B} \tau_2$ to mean $w(\sat(F, \tau_1)
  \setminus B) \leq w(\sat(F, \tau_2) \setminus B)$ both for the
  definition of $\mathtt{Tab}$ and for Procedure~3, we are able to solve
  the more general problem weighted \textsc{MaxSAT} in the same way.
  
  For the problem \textsc{\#SAT}, we care only about assignments
  satisfying  all the clauses of $F$, and we want to decide the
  number of distinct assignments doing so. This requires a few alterations.
  Firstly, alter the definition of the contents of
  $\mathtt{Tab}_v(C, C')$ in Constraint (\ref{eq}) to be the number of assignments $\tau$
  over $\var(\delta(v))$ where $\sat(F_v,\tau) = C$ and
  $\cla(\delta(v)) \setminus C' \subseteq \sat(F, \tau)$. Secondly, when computing
  $\mathtt{Tab}_l$ for the leaves $l$ of $T$, we set
  each of the entries of $\mathtt{Tab}_l$ to either zero, one, or two,
  according to the definition.  
  Thirdly, we alter the algorithm to compute $\mathtt{Tab}_v$
  (Procedure~3) for inner nodes. We initialize $\mathtt{Tab_v}(C,C')$ to be
  zero  at the start of the algorithm, and
  substitute lines 6, 7 and 8 of Procedure~3 by the following line which increases the table value by 
  the product of the table values at the children
    \begin{quote}
    \begin{tabbing}
    $\mathtt{Tab_v}(C_v, {C_{\compl v}}) \leftarrow \mathtt{Tab_v}(C_v, {C_{\compl v}})+ \mathtt{Tab_{c_1}}(C_{c_1}, C_{\compl {c_1}}) \cdot \mathtt{Tab_{c_2}}(C_{c_2}, C_{\compl {c_2}})$\\
    \end{tabbing}
  \end{quote}
  This will satisfy our new constraint of $\mathtt{Tab}_v$ for
  internal nodes $v$ of $T$.   The
  value of $\mathtt{Tab}_r(\emptyset, \emptyset)$ at the root $r$ of $T$ will be exactly the
  number of distinct assignments satisfying all clauses of $F$.
\end{qedproof}

The bottleneck giving the cubic factor $k^3$ in the runtime of
Theorem~\ref{theorem:SolvingSATonBranchDec} is the number triples in 
$\mathtt{PS}(F_{\compl v}) \times
\mathtt{PS}(F_{c_1}) \times \mathtt{PS}(F_{c_2})$ for any node $v$ with
children $c_1$ and $c_2$. When $(T, \delta)$ is a linear branch
decomposition, it is always the case that either $c_1$ or $c_2$ is a leaf of $T$. 
In this case either $|\mathtt{PS}(F_{c_1})|$ or
$|\mathtt{PS}(F_{c_2})|$ is a constant. Therefore, for linear branch decompositions
$\mathtt{PS}(F_{\compl v}) \times \mathtt{PS}(F_{c_1}) \times
\mathtt{PS}(F_{c_2})$ will contain no more than $\bigoh(k^2)$ triples.
Thus we can reduce the runtime of the algorithm by a
factor of $k$. 

\begin{theorem}\label{theorem:SolvingSATonLinearBranchDec}
  Given a formula $F$ over $n$ variables and $m$ clauses and of size $s$, and a linear branch decomposition
  $(T, \delta)$ of $F$ of $\mathtt{ps}$-width $k$, we solve
  \textsc{\#SAT}, \textsc{MaxSAT}, and weighted \textsc{MaxSAT} in
  time $\bigoh(k^2s(m + n))$.
\end{theorem}

\section{CNF formulas of polynomial $\mathtt{ps}$-width}

In this section we investigate classes of CNF formulas having decompositions with $\mathtt{ps}$-width polynomially bounded in formula size $s$.
In particular, we show that this holds whenever the incidence graph of the formula has constant MIM-width (maximum induced matching-width).
We also show that a large class of bipartite graphs, using what we call bigraph bipartizations, have constant MIM-width.

Let us start by defining bigraph bipartizations. 
For a graph $G$ and subset of vertices $A \subseteq V(G)$ the bipartite graph $G[A, \compl{A}]$ is the subgraph of $G$
containing all edges of $G$ with exactly one endpoint in $A$. We call $G[A, \compl{A}]$ a bigraph bipartization of $G$, note that 
$G$ has a bigraph bipartization for each subset of vertices.
For a graph class $X$ define the class of $X$ bigraphs as the bipartite graphs $H$ for which there exists $G \in X$ such that
$H$ is isomorphic to a bigraph bipartization of $G$.
For example, $H$ is an interval bigraph if there is some interval graph $G$ and some $A \subseteq V(G)$ with $H$ isomorphic to $G[A, \compl{A}]$.

To establish the connection to MIM-width we need to look at induced matchings in the incidence graph of a formula.
  The incidence graph of a formula $F$ is the bipartite graph $I(F)$ having a vertex for each clause and variable, with variable $x$ adjacent to any clause $C$ in which it occurs.
  An induced matching in a graph is a subset $M$ of edges with the property that any edge of the graph is incident to at most one edge in $M$. In other words, for any 3 vertices $a,b,c$, if $ab$ is an edge in $M$ and $bc$ is an edge then there does not exist an edge $cd$ in $M$.
 The number of edges in $M$ is called the size of the induced matching.
  The following result provides an upper
bound on the $\mathtt{ps}$-value of a formula in terms of the maximum size of an induced matching of its incidence graph.

 \begin{lemma}\label{lem:ps_leq_mimw}
  Let $F$ be a CNF formula and let $k$ be the maximum size of an induced matching in $I(F)$. 
  We then have $|\mathtt{PS}(F)| \leq |\cla(F)|^k$.
 \end{lemma}

 \begin{qedproof}
  Let $\mathcal{C} \in \mathtt{PS}(F)$ and $\mathcal{C}_{f} = \cla(F) \setminus \mathcal{C}$. Thus, there exists a complete assignment $\tau$  
such that the clauses not satisfied by $\tau$ are $\mathcal{C}_{f} = \cla(F) \setminus \sat(F,\tau)$.
Since every variable in $\var(F)$ appears in some clause of $F$ this means that $\tau\vert_{\var(\mathcal{C}_{f})}$ is the unique assignment of the variables in $\var(\mathcal{C}_{f})$ which
do not satisfy any clause of $\mathcal{C}_{f}$.
  Let $\mathcal{C}_{f}^{'} \subseteq \mathcal{C}_{f}$ be an inclusion minimal set such that $\var(\mathcal{C}_{f}) = \var(\mathcal{C}_{f}^{'})$, hence
$\tau\vert_{\var(\mathcal{C}_{f})}$ is also the unique assignment of the variables in $\var(\mathcal{C}_{f})$ which
do not satisfy any clause of $\mathcal{C}_{f}^{'}$.
An upper bound on the number of different such minimal $\mathcal{C}_{f}^{'}$, over all $\mathcal{C} \in \mathtt{PS}(F)$, will give an upper bound on $|\mathtt{PS}(F)|$.
 For every $C \in \mathcal{C}_{f}^{'}$ there is a variable $v_C$ appearing in $C$ and no other clause of $\mathcal{C}_{f}^{'}$, 
otherwise $\mathcal{C}_{f}^{'}$ would not be minimal. Note that we have an induced matching $M$ of $I(F)$ containing
all such edges $v_C,C$.
  By assumption, the induced matching $M$ can have at most $k$ edges and hence $|\mathcal{C}_{f}^{'}| \leq k$.
  There are at most $|\cla(F)|^k$ sets of at most $k$ clauses and the lemma follows. 
\end{qedproof}

In order to lift this result on the $\mathtt{ps}$-value of $F$, i.e $|\mathtt{PS}(F)|$, to the $\mathtt{ps}$-width of $F$, we use MIM-width of the incidence graph $I(F)$, which is defined using branch decompositions of graphs. 
A branch decomposition of the formula $F$, as defined in Section 2, can also be seen as a branch decomposition of the incidence graph $I(F)$. Nevertheless, for completeness, we formally define branch decompositions of graphs and MIM-width.

A branch decomposition of a graph $G$ is a pair $(T,\delta)$ where $T$ is a rooted binary tree  and $\delta$ a bijection between the leaf set of $T$ and the vertex set of $G$.
For a node $w$ of $T$ let the subset of $V(G)$ in bijection $\delta$ with the leaves of the subtree of $T$ rooted at $w$ be denoted by $V_w$. 
We say the decomposition defines the cut $(V_w,\compl{V_w})$.
The MIM-value of a cut $(V_w,\compl{V_w})$ is the size of a maximum induced matching of $G[V_w,\compl{V_w}]$.
The MIM-width of $(T,\delta)$ is the maximum MIM-value over all cuts $(V_w,\compl{V_w})$ defined by a node $w$ of $T$.
The MIM-width of graph $G$, denoted $mimw(G)$, is the minimum MIM-width over all branch decompositions $(T,\delta)$ of $G$.
As before a \emph{linear branch decomposition} is a branch decomposition where inner nodes of the underlying tree induces a path.

We now give an upper bound on the $\mathtt{ps}$-value of a formula in terms of the MIM-width of any graph $G$ such that the incidence graph of the formula is a bigraph bipartization of $G$.

\begin{theorem}\label{thm:biparte}
  Let $F$ be a CNF formula of $m$ clauses, $G$ a graph, and $(T, \delta_G)$ a
  (linear) branch decomposition of $G$ of MIM-width $k$. If for a
  subset $A \subseteq V(G)$ the graph $G[A, \compl A]$ is isomorphic
  to $I(F)$, then we can in linear time produce a (linear) branch
  decomposition $(T, \delta_F)$ of $F$ having $\mathtt{ps}$-width at
  most $m^k$.
\end{theorem}

\begin{qedproof}
  Since each variable and clause in $F$ has a corresponding node in
  $I(F)$, and each node in $I(F)$ has a corresponding node in $G$, by
  defining $\delta_F$ to be the function mapping each leaf $l$ of $T$
  to the variable or clause in $F$ corresponding to the node
  $\delta_G(l)$, $(T, \delta_T)$ is going to be a branch decomposition
  of $F$.
  For any cut $(A,\compl{A})$
  induced by a node of $(T,\delta_F)$, let $C \subseteq \cla(F)$ be the
  clauses corresponding to vertices in $A$ and $X \subseteq \var(F)$
  the variables corresponding to vertices in $A$.  The cut $S = C \cup
  X$ of $F$ defines the two formulas $F_{C,\compl{X}}$ and
  $F_{\compl{C},X}$, and it holds that $I(F_{C,\compl{X}})$ and $I(F_{\compl{C},X})$ are 
  induced subgraphs of $G[A,\compl{A}]$ and hence by
  Lemma~\ref{lem:ps_leq_mimw}, we have $|\mathtt{PS}(F_{C,\compl{X}})|
  \leq |\cla(F)|^{\mathtt{mim}(A)}$, and likewise we have $|\mathtt{PS}(F_{\compl{C},X})|
  \leq |\cla(F)|^{\mathtt{mim}(A)}$. Since the $\mathtt{ps}$-width of the
  decomposition is the maximum $\mathtt{ps}$-value of each cut, the
  theorem follows.
\end{qedproof}

Note that by taking $G=I(F)$ and $A=\cla(F)$ and letting $(T,
\delta_G)$ be a branch decomposition of $G$ of minimum MIM-width, we
get the following weaker result.

\begin{corollary}
  For any CNF formula $F$ over $m$ clauses, the $\mathtt{ps}$-width of $F$ is no larger than
  $m^{\mathtt{mimw}(I(F))}$.
\end{corollary}

In his thesis, Vatshelle~\cite{Vatshelle} shows that MIM-width of any
graph $G$ is at most the clique-width of $G$. Furthermore, the
clique-width has been shown by
Courcelle~\cite{DBLP:journals/dm/Courcelle04} to be at most twice the
symmetric clique-width. Thus, we can conclude that MIM-width is
bounded on any graph class with a bound on the symmetric clique-width,
in accordance with Figure~\ref{fig:6}.

 Many classes of graphs have intersection models, meaning that they can be represented as intersection graphs of certain objects, i.e.\ each vertex is associated with an object and two
vertices are adjacent iff their objects intersect.
 The objects used to define intersection graphs usually consist of geometrical objects such as lines, circles or polygons.
 Many well known classes of intersection graphs have constant MIM-width, as in the following which lists only a subset of the classes proven to have such bounds in \cite{DBLP:journals/tcs/BelmonteV13,Vatshelle}.

\begin{theorem}[\cite{DBLP:journals/tcs/BelmonteV13,Vatshelle}] \label{VatsBelm}
 Let $G$ be a graph. If $G$ is a:\\
 \indent~~~interval graph then $\mathtt{mimw}(G) \leq 1$.\\
 \indent~~~circular arc graph then $\mathtt{mimw}(G) \leq 2$.\\
 \indent~~~$k$-trapezoid graph then $\mathtt{mimw}(G) \leq k$.\\
Moreover there exist linear decompositions satisfying the bound.
\end{theorem}

Let us briefly mention the definition of these graph classes.
A graph is an interval graph if it has an intersection model consisting of intervals of the real line.
A graph is a circular arc graph if it has an intersection model consisting of arcs of a circle.
To build a $k$-trapezoid we start with $k$ parallel line segments $(s_1,e_1),(s_2,e_2),...,(s_k,e_k)$ and 
add two non-intersecting paths $s$ and $e$ by joining $s_i$ to $s_{i+1}$ and $e_i$ to $e_{i+1}$ respectively by straight lines
for each $i \in \{1,...,k-1\}$. The polygon defined by $s$ and $e$ and the two line segments $(s_1,e_1),(s_k,e_k)$  forms a $k$-trapezoid.
A graph is a $k$-trapezoid graph if it has an intersection model consisting of $k$-trapezoids.
See \cite{BraBanSpi99} for information about graph classes and their containment relations.
Combining Theorems \ref{thm:biparte} and \ref{VatsBelm} we get the following.

\begin{corollary}
  Let $F$ be a CNF formula containing $m$ clauses. If $I(F)$ is a:\\
  \indent~~~interval bigraph then $\mathtt{psw}(F) \leq m$.\\
  \indent~~~circular arc bigraph then $\mathtt{psw}(F) \leq m^2$.\\
  \indent~~~$k$-trapezoid bigraph then $\mathtt{psw}(F) \leq m^k$.\\
  Moreover there exist linear decompositions satisfying the bound.
\end{corollary}

\section{Interval bigraphs and formulas having interval orders}

We will in this section show one class of formulas where we can find linear branch decompositions
having $\mathtt{ps}$-width $\bigoh(|\cla(F)|)$. Let us recall the definition of interval ordering.
A CNF formula $F$ has an interval ordering if there exists a linear ordering of variables and clauses
such that for any variable $x$ occurring in clause $C$, if $x$ appears before $C$
then any variable between them also occurs in $C$, and if $C$ appears before $x$
then $x$ occurs also in any clause between them.
By a result of Hell and Huang ~\cite{DBLP:journals/jgt/HellH04} it follows that a formula $F$ has an interval
ordering if and only if $I(F)$ is a interval bigraph.

\begin{theorem} \label{thm:intbigraph_algorithm} Given a CNF formula
  $F$ over $n$ variables and $m$ clauses and of size $s$, we can in time $\bigoh((m + n)s)$
  decide if $F$ has an interval ordering (yes iff $I(F)$ is an
  interval bigraph), and if yes we solve \textsc{\#SAT} and weighted
  \textsc{MaxSAT} with a runtime of
  $\bigoh(m^2(m + n)s)$.
\end{theorem}
\begin{qedproof}
  Using the characterization of ~\cite{DBLP:journals/jgt/HellH04} and
  the algorithm of \cite{DBLP:journals/corr/abs-1211-2662} we can in
  time $\bigoh((m + n)s)$ decide if $F$ has an
  interval ordering and if yes, then we find it.  From this interval
  ordering we build an interval graph $G$ such that $I(F)$ is a
  bigraph bipartization of $G$, and construct a linear branch
  decomposition of $G$ having MIM-width
  $1$~\cite{DBLP:journals/tcs/BelmonteV13}.  From such a linear branch
  decomposition we get from Theorem~\ref{thm:biparte} that we can
  construct another linear branch decomposition of $F$ having
  $\mathtt{ps}$-width $\bigoh(m)$.  We then run the algorithm
  of Theorem~\ref{theorem:SolvingSATonLinearBranchDec}.
\end{qedproof}

\section{Conclusion}

In this paper we have proposed a structural parameter of CNF formulas, called $\mathtt{ps}$-width
or perfectly-satisfiable-width.
We showed that weighted {\sc MaxSAT} and {\sc \#SAT} can be solved in polynomial time
on formulas  given with a decomposition of polynomially bounded $\mathtt{ps}$-width.
Using the concept of interval bigraphs we also showed a polynomial time algorithm that actually finds such a decomposition, for formulas having an interval ordering. 

Could one devise such an algorithm also for the larger class of circular arc bigraphs, or maybe even for the even larger class of $k$-trapezoid bigraphs? In other words, is the problem of recognizing if a bipartite input graph is a circular arc bigraph, or a $k$-trapezoid bigraph, polynomial-time solvable?

It could be interesting to give an algorithm solving {\sc MaxSAT} and/or {\sc \#SAT} directly on the interval ordering of a formula, rather than using the more general notion of $\mathtt{ps}$-width as in this paper. Maybe such an algorithm could be of practical use?

Also of practical interest would be to design a heuristic algorithm which given a formula finds a decomposition of relatively low $\mathtt{ps}$-width, as has been done for boolean-width in \cite{HvidevoldSTV11}.

Finally, we hope the essential combinatorial result enabling the improvements in this paper, Lemma \ref{lem:ps_leq_mimw}, may have other uses as well.

\bibliography{sat}

\end{document}